\documentclass[twocolumn,preprintnumbers,amsmath,amssymb]{revtex4}
\usepackage{amssymb}
\usepackage{latexsym}
\usepackage{epsfig}
\usepackage{color}
\usepackage{float}
\begin{document}

\title{Thermodynamic approach to holographic dark energy and the R\'{e}nyi entropy}

\author{H. Moradpour$^1$\footnote{h.moradpour@riaam.ac.ir}, S. A. Moosavi$^1$, I. P.
Lobo$^{2}$\footnote{iarley\_lobo@fisica.ufpb.br}, J. P. Morais Gra\c ca$^{2}$\footnote{jpmorais@gmail.com}, A.
Jawad$^3$\footnote{abduljawad@ciitlahore.edu.pk}, I. G.
Salako$^{4}$\footnote{inessalako@gmail.com}}
\address{$^1$ Research Institute for Astronomy and Astrophysics of Maragha
(RIAAM), P.O. Box 55134-441, Maragha, Iran\\
$^2$ Departamento de F\'{i}sica, Universidade Federal da
Para\'{i}ba, Caixa Postal 5008, CEP 58051-970, Jo\~{a}o Pessoa,
PB, Brazil\\
$^{3}$ Department of Mathematics, COMSATS Institute of Information Technology, Lahore-54000, Pakistan\\
$^4$ Institut de Math\'ematiques et de Sciences Physiques (IMSP)
01 BP 613 Porto-Novo, B\'enin}

\begin{abstract}
Using the first law of thermodynamics, we propose a relation between
the system entropy ($S$) and its IR ($L$) and UV ($\Lambda$)
cutoffs. In addition, applying this relation to the apparent horizon
of flat FRW universe, whose entropy meets the R\'{e}nyi entropy, a
new holographic dark energy model is addressed. Thereinafter, the
evolution of the flat FRW universe, filled by a pressureless source
and the obtained dark energy candidate, is studied. In our model,
there is no mutual interaction between the cosmos sectors. We find
out that the obtained model is theoretically powerful to explain the
current accelerated phase of the universe. This result emphasizes
that the generalized entropy formalism is suitable for describing
systems including the long-range interactions such as gravity.
\end{abstract}
\maketitle
\section{Introduction\label{Intr}}

The nature of current accelerated universe
\cite{Riess0,Riess,Riess1,Riess2,Riess3,COL2001,COL20011,COL20012,
COL20013,HAN2000,HAN20001,HAN20002}, related to an unknown source
called dark energy (DE), is one of the unsolved physical mysteries
\cite{mat1,mat2,mat3,Roos}. In order to solve it, modified
theories of gravity have been proposed which describe it in terms
of various geometrical effects
\cite{meeq}. On the other hand,
introducing the new types of matter or diverse state equations constitutes
some other theoretical attempts to solve the DE
problem
\cite{DEC3,DEC6,DEC8,DEC12,DEC13,DEC27,DEC28,DEC372,DEC39,Rev1,Rev2}. It has been also
argued that there are deep connections between DE, the horizon entropy
and the thermodynamics laws \cite{mswr,em,DEC37,ijtpmr,pav}.

It seems that systems including the long-range interactions, such as
gravity, are more in agreement with the generalized entropy
formalisms based on the power-law distribution of probabilities
\cite{pla,nn1,nn2,nn3,non0,non1,abe,fon,smm,5,non2}. Recently, two generalized entropies including the
R\'{e}nyi and Tsallis entropies
\cite{pla,nn1,nn2,nn3,non0,non1,abe,fon}, have been used in order
to study various gravitational and cosmological phenomena
\cite{non2,non16,non19,non20,non21,non22,non13,non4,non5,non6,non7,5,smm}.
It has been shown in various ways that the generalized entropy
formulation can provide suitable descriptions for DE, also
motivating us to consider generalized entropies as the horizon
entropy instead of the Bekenstein entropy
\cite{5,non4,non5,non6,non7,non13,non19,non20,
non21,smm}. In fact, the Bekenstein entropy is also obtainable by
applying the Tsallis statistics to the systems including gravity
\cite{5,non2,non22,non19,non20,non21}.

To reconcile the breakdown of quantum field theory in large scale
with the success of effective field theory, Cohen et al,
\cite{HDE} proposed a new relation between the system entropy
($S$) together with the IR ($L$) and UV ($\Lambda$) cutoffs which
finally lead to \cite{HDE5,HDE17}

\begin{eqnarray}\label{coh}
\rho_\Lambda\propto\frac{S}{L^4},
\end{eqnarray}

\noindent where $\rho_\Lambda\sim\Lambda^4$ is the vacuum energy
density. Applying this hypothesis to cosmological setup, authors got
a model for DE called Holographic Dark Energy (HDE), in which
$\rho_\Lambda$ plays the role of the energy density of DE
($\rho_d$), \cite{HDE01,HDE1,HDE2}. One problem with the original
model (OHDE) is that if one uses the Hubble radius as the IR cutoff,
and considers the Bekenstein entropy, then both dark matter (DM) and
OHDE are scaled with the same function of scale factor
\cite{HDE2,HDE3}. Although this problem may be solved by introducing
new cutoffs \cite{HDE2,HDE3}, such cutoffs cannot always lead to
stable models for DE whenever it is dominant in cosmos \cite{stab}.
More studies on HDE and its various features can be found in
Refs.~\cite{HDE6,RevH,wang}.

Recently, using Eq.~(\ref{coh}) and a generalized entropy, called
the Sharma-Mittal measure \cite{pla}, a new model of HDE has been
proposed and studied \cite{smm}. In this new model (SMHDE), Hubble
horizon is the IR cutoff, and there is no interaction between the
cosmos components \cite{smm}. SMHDE is compatible with the universe
expansion history, and it is stable whenever it is dominant in
cosmos \cite{smm}. Hence, this model suffers from less weakness
compared to the OHDE corrected by considering other cutoffs
\cite{smm,stab}. The obtained behavior of SMHDE also
motivates us to farther study the HDE hypothesis in other generalized
entropy formalisms. Such analysis may help us to become more
familiar with thermodynamics and the statistical aspects of
spacetime and gravity \cite{smm,non20,5}.

It is also useful to remind here that the apparent horizon is a
proper causal boundary for cosmos in agreement with the
thermodynamics laws and thus the energy-momentum conservation law
\cite{cons,cons1,cons2,cons3,Cai2,CaiKimt}. Moreover, in flat FRW
universe, Friedmann equations indicate that whenever DE is dominant
in cosmos, its energy density will scale with $H^2$ \cite{Roos}.
Therefore, from a thermodynamic point of view, a HDE model in flat
FRW universe, for which the Hubble radius and the radius of apparent
horizon ($1/H$) are the same, will be more compatible with the
thermodynamics laws, if it can provide a proper description for
the universe by using the Hubble radius as its IR cutoff.
It is worthwhile mentioning that the first law of thermodynamics
(FLT) helps us in relating the equation of state of HDE to that of
black holes leading to a model for the current accelerated universe
\cite{my}. FLT is also employed in building HDE models based on
energy density of vacuum entanglement \cite{RevH,E1}. As a result,
one can argue that we should use the thermodynamics laws in order to
build a relation between $\Lambda^4$, the IR cutoff and the horizon
entropy to get a model for HDE.

Here, considering FLT, a relation between the system entropy ($S$)
together with the IR ($L$) and UV ($\Lambda$) cutoffs will be
proposed. In addition, using the Hubble radius as the IR cutoff, and
employing a $Q$-generalized entropy, proposed by R\'{e}nyi
\cite{non0}, we are going to introduce a new HDE model in flat FRW
universe. In order to achieve our goal, the paper is organized as
follows. Our thermodynamic version of Eq.~(\ref{coh}) has been
introduced in the next section. In section~(\textmd{III}), a brief
review on the R\'{e}nyi and Tsallis entropy is given, and
then, we introduce our new model of HDE. The behavior of the model
whenever there is no interaction between the cosmos sectors is
studied in Sec.~($\textmd{IV}$). The last section is devoted to a
summary and concluding remarks. We also work in the unit of
$c=\hbar=G=k_B=1$, where $k_B$ denotes the Boltzmann constant.

\section{A thermodynamic version for HDE}

Bearing the Cai-Kim
temperature ($T=\frac{1}{2\pi L}$) in mind \cite{CaiKimt}, for a system with IR cutoff $L$, one can
reach Eq.~(\ref{coh}) by considering the $E_d\sim\rho_d V\simeq
E_T\propto TS$ assumption, in which $\rho_d\equiv\rho_\Lambda\sim\Lambda^4$, \cite{HDE5,HDE17}. In fact, since DE is dominant in the current cosmos, the $E_d\sim\rho_d V\simeq
E_T\propto TS$ assumption is acceptable. Here, $V=\frac{4\pi L^3}{3}$ is the
aerial volume of FRW spacetime, $S$ is the
horizon entropy (represents the total entropy of system), $E_T$ also denotes the total energy content
of cosmos \cite{HDE5,HDE17}. Moreover, $E_d$ represents the energy content of the DE candidate (vacuum energy) \cite{HDE5,HDE17}. Recently, the above assumption has also been used in
order to provide a thermodynamic description for HDE, and also a
relation between holographic minimal information density and the
de Broglie's wavelength \cite{HDET}. Although Eq.~(\ref{coh}) is compatible with the dimensional
analysis \cite{HDE2}, but since only in flat FRW universe the aerial volume is the same as the actual volume ($L=\frac{1}{H}$), the
above argument, and thus Eq.~(\ref{coh}) are more reliable in the flat
FRW universe \cite{vv1,vv2,HDE}. Hence, since the non-flat FRW universe has not completely been rejected \cite{Roos}, it is better to modify the $E_d\sim\rho_d V\simeq
E_T\propto TS$ assumption in a more consistent way with the non-flat cases.

For the first time, using the $dE_T\equiv d\mathcal{Q}$ assumption
in order to find the energy flux ($d\mathcal{Q}$) seen by an
accelerated observer inside horizon, and applying the Calusius
relation to the horizon, Jacobson wrote the Einstein field
equations in the static spacetimes as a thermodynamic equation of
state \cite{j1}. The generalizations of his idea to the
cosmological setups suggest that the
$dE_T=TdS$ relation can be considered as the first law of thermodynamics
for the cosmological horizon \cite{cons3,j2}.

The above arguments motivate us to introduce $dE_d=\rho_d dV\propto
dE_T=TdS$ for building a thermodynamic consistent relation between
the UV cutoff ($\Lambda^4\sim\rho_\Lambda\equiv\rho_d)$ and the IR
cutoff. In order to check this assumption, consider a flat FRW
universe for which $S=\frac{\pi}{H^2}$, $T=\frac{H}{2\pi}$ and
$V=\frac{4\pi}{3H^3}$, where $H$ is the Hubble parameter. By using
$\rho_d dV\propto TdS$, we easily reach
$\rho_d\propto\frac{H^2}{4\pi}$ which is nothing but OHDE for which
Hubble horizon is considered as the IR cutoff
\cite{HDE01,HDE1,HDE2}. In fact, our thermodynamic reformulation of
Eq.~(\ref{coh}) claims that the changes in the energy (entropy) of
system can not be more than that of the black hole of the same size.
In other words, we have $dE_d\leq dE_T$ in every infinitesimal
interval leading to $E_d=\int dE_d\leq E_T=\int dE_T$, a result in
agreement with the HDE hypothesis
\cite{HDE,HDE5,HDE17,HDE01,HDE1,HDE2,HDE3,stab,HDE6}. Therefore, in
our setup, the final amount of the system energy (entropy) can not
also be more than that of its same size black hole.


\section{R\'{e}nyi entropy and HDE: general remarks}

For a system consisting of $W$ discrete states, Tsallis entropy is
defined as \cite{non1}

\begin{eqnarray}\label{reyn02}
S_T=\frac{1}{1-Q}\sum_{i=1}^{W}(P_i^Q-P_i),
\end{eqnarray}

\noindent in which $P_i$ denotes the ordinary probability of
accessing state $i$, and $Q$ is a real parameter which may be
originated from the non-extensive features of system such as the long
range nature of gravity \cite{5,smm,non1,nn3}. In this formalism, two
probabilistically independent systems obey the non-additive
composition rule \cite{non1,abe}

\begin{eqnarray}\label{nona1}
S_{12}=S_1+S_2+\delta S_1 S_2,
\end{eqnarray}

\noindent where $\delta\equiv1-Q$. Thus, even for
probabilistically independent systems, although it is a non-additive
entropy, unless we have $\delta=0$ \cite{non1}, it is also not always
non-extensive \cite{nn3}. In fact, differences between
non-additivity and non-extensivity are very delicate
\cite{nn1,nn2,nn3}, and the concept of non-extensivity is much
more complex than that of the non-additivity \cite{nn3}. For
example, the famous Bekenstein entropy is non-extensive and
non-additive simultaneously \cite{non2,non22}.

In addition, there is another $Q$-generalized entropy definition as

\begin{eqnarray}\label{reyn01}
\mathcal{S}=\frac{1}{1-Q}\ln\sum_{i=1}^{W} P_i^Q,
\end{eqnarray}

\noindent which returns to A. R\'{e}nyi \cite{non0}. 
One can combine Eqs.~(\ref{reyn01}) and~(\ref{reyn02}) with each other
to reach \cite{non19,non20,non21}

\begin{eqnarray}\label{reyn1}
\mathcal{S}=\frac{1}{\delta}\ln(1+\delta S_T),
\end{eqnarray}

\noindent as the R\'{e}nyi entropy content of system. Thus, based on Eq.~(\ref{reyn1}), $\mathcal{S}$
will remain additive as long as Eq.~(\ref{nona1}) is obtained by
$S_T$ \cite{pla,non22,nn2}.
%
%
%
The above $Q$-generalized entropies can be used whenever systems are described better by
using the power law distributions $P_i^Q$, also called $Q$-distribution, instead of the ordinary probability distribution
$P_i$ \cite{non22,nn2}. Systems including long-range interactions, such as gravity, are primary candidates for using $Q$-distribution \cite{pla,nn1,nn2,nn3,non0,non1,abe,fon,smm,5,non2}.

It is also useful to remind here that a $Q$-distribution-based
description of the universe can theoretically describe the current
accelerated universe
\cite{non4,non5,non6,non7,non13,non19,non20,non21,smm} which more motivates
us to assign various $Q$-generalized entropies to the cosmological
horizons. Recently, it has been argued that the Bekenstein entropy
($S=\frac{A}{4}$) is in fact a Tsallis
entropy \cite{5,abe,non2,non19,non20,non21,non22}
leading to

\begin{eqnarray}\label{renyif}
\mathcal{S}=\frac{1}{\delta}\ln(1+\frac{\delta}{4}A),
\end{eqnarray}

\noindent for the R\'{e}nyi entropy content of system
\cite{non2,non22,non19,non20,non21}.

\subsection*{R\'{e}nyi Holographic Dark Energy (RHDE)}

Here, we only focus on the flat FRW universe indicated by the WMAP
data \cite{Roos}. In our model, the vacuum energy density plays the
role of DE meaning that we have
$\Lambda^4\sim\rho_\Lambda\equiv\rho_d$. Now, using the $\rho_d
dV\propto TdS$ assumption and Eq.~(\ref{renyif}), one can get RHDE
as

\begin{eqnarray}\label{hded}
\rho_d=\frac{3C^2
H^2}{8\pi(1+\frac{\delta\pi}{H^2})},
\end{eqnarray}

\noindent where $C^2$ is a numerical constant as usual. In
order to obtain this equation, we also used the $T=\frac{H}{2\pi}$
and $A=\frac{4\pi}{H^2}=4\pi(\frac{3V}{4\pi})^{\frac{2}{3}}$
relations, valid in a flat FRW universe \cite{CaiKimt}. It is apparent that in the
absence of $\delta$, we have $\rho_d=\frac{3C^2H^2}{8\pi}$ in full
agreement with OHDE \cite{HDE01,HDE1,HDE2}.

In order to obtain the corresponding pressure, we assume that RHDE
obeys ordinary energy-momentum conservation law in the FRW universe
with scale factor $a$, and thus

\begin{eqnarray}\label{emc1}
\dot{\rho}_d+3H(\rho_d+p_d)=0,
\end{eqnarray}

\noindent in which dot denotes derivative with respect to time. It
finally leads to

\begin{eqnarray}\label{hdep}
p_d&=&-\frac{\rho_d^\prime\dot{H}}{3H}-\rho_d,
\end{eqnarray}

\noindent where prime denotes derivative with respect to $H$, for
the pressure of RHDE.


\section{Universe evolution}

For a flat FRW universe filled by a pressureless source with
energy density $\rho_m$ and RHDE, the Friedmann equations are
written as

\begin{eqnarray}\label{fe1}
&&H^2=\frac{8\pi}{3}(\rho_m+\rho_d),\\
&&H^2+\frac{2}{3}\dot{H}=\frac{-8\pi}{3}(p_d).\nonumber
\end{eqnarray}

\noindent Let us define the density parameters corresponding
to the $\rho_m$ and $\rho_d$ sources as
$\Gamma_m\equiv\frac{\rho_m}{\rho_c}$ and
$\Omega_D\equiv\frac{\rho_d}{\rho_c}$, respectively, where
$\rho_c(\equiv\frac{3H^2}{8\pi})$ is called critical density. Now,
inserting these definitions in the first line of Eq.~(\ref{fe1}),
one easily finds

\begin{eqnarray}\label{fe2}
1=\Gamma_m+\Omega_D.
\end{eqnarray}

\noindent Here, since DE candidate obeys
Eq.~(\ref{emc1}), there is no interaction between matter source and
RHDE meaning that we have

\begin{eqnarray}\label{dmd}
\dot{\rho}_m+3H\rho_m=0\Rightarrow \rho_m=\rho_0 a^{-3},
\end{eqnarray}

\noindent in which $\rho_0$ is the integration constant equal to the
current value of the energy density of pressureless component
($\rho_m$). Now, defining $H(z)=E(z)H_0$, bearing the $1+z=a^{-1}$
relation in mind, where $z$ and $H_0$ denote redshift and the
current value of the Hubble parameter, respectively, and combining
Eqs.~(\ref{dmd}) and~(\ref{hded}) with either the first
equation of~(\ref{fe1}) or Eq.~(\ref{fe2}), one easily finds that

\begin{eqnarray}\label{z}
E^2(z)=\Omega_m(1+z)^3+\frac{(1+\frac{\delta\pi}{H_0^2})(1-\Omega_m)}{1+\frac{\delta\pi}{E^2(z)H_0^2}}E^2(z),
\end{eqnarray}

\noindent where
$\Omega_m\equiv\Gamma_m(z=0)=\frac{\rho_0}{\frac{3H_0^2}{8\pi}}$
denotes the current value ($z=0$) of the density parameter of
$\rho_m$. In obtaining this equation, we assumed that for $z=0$ we
have $E(z)=1$ leading to
$C^2=(1+\frac{\delta\pi}{H_0^2})(1-\Omega_m)$, which can finally be
combined with Eq.~(\ref{hded}) to get

\begin{eqnarray}\label{hded1}
\rho_d=\frac{3(1+\frac{\delta\pi}{H_0^2})(1-\Omega_m)
H^2}{8\pi(1+\frac{\delta\pi}{H^2})},
\end{eqnarray}

\noindent for energy density. Deceleration parameter is also
evaluated as

\begin{eqnarray}\label{decp2}
q\equiv-1-\frac{\dot{H}}{H^2}=-1+\frac{1+z}{E(z)}\frac{dE(z)}{dz}.
\end{eqnarray}

\noindent Moreover, if we characterize the total state equation of
cosmos as $w\equiv\frac{p_d}{\rho_d+\rho_m}$, then using
Eqs.~(\ref{fe1}) and~(\ref{decp2}), we can find

\begin{eqnarray}\label{Eos}
w=\frac{2}{3}(q-\frac{1}{2}).
\end{eqnarray}

\noindent Bearing the definitions of $\Omega_D$ and
$\Gamma_m$ in mind, one can use Eq.~(\ref{z}) to reach at

\begin{eqnarray}\label{dp1}
\Omega_D&=&\frac{(1+\frac{\delta\pi}{H_0^2})(1-\Omega_m)}{1+\frac{\delta\pi}{E^2(z)H_0^2}},
\end{eqnarray}

\noindent for the density parameter of RHDE. In order to
investigate the stability of RHDE, we need to study the evolution of
square of the sound speed evaluated as

\begin{equation}\label{1}
v_{s}^{2}=\frac{dp_d}{d\rho_d}=\frac{\frac{dp_d}{dH}}{\frac{d\rho_d}{dH}}.
\end{equation}

\noindent One can use the second Friedmann
equation~(\ref{fe1}) to find $\dot{H}=-4\pi p_d-\frac{3H^2}{2}$.
Now, employing this result along with Eqs.~(\ref{hdep})
and~(\ref{hded1}), we reach at



\begin{eqnarray}\label{pf}
&&p_d=\frac{\frac{\delta\pi\rho^2_d}{\alpha H^4}}{\frac{8\pi\rho_d}{3H^2}(\frac{\delta\pi\rho_d}{\alpha H^4}-1)-1},\\
&&\alpha=\frac{3}{8\pi}(1+\frac{\delta\pi}{H_0^2})(1-\Omega_m).\nonumber
\end{eqnarray}

\noindent This result together with Eq.~(\ref{hded1}) can be
used to evaluate $v_{s}^{2}$ as

\begin{eqnarray}\label{vf}
v_{s}^{2}&=&\frac{1-(1+\frac{\delta\pi}{H_0^2})(1-\Omega_m)+\frac{\delta\pi}{H^2}}{[1-\frac{(1+\frac{\delta\pi}{H_0^2})(1-\Omega_m)(1+2\frac{\delta\pi}{H^2})}{(1+\frac{\delta\pi}{H^2})^2}]^2}
\bigg[-H^2\\
&+&\frac{1+3\frac{\delta\pi}{H^2}}{(1+\frac{\delta\pi}{H^2})^2}
-\frac{\frac{\delta\pi}{H^2}(3+\frac{\delta\pi}{H^2})}{(1+2\frac{\delta\pi}{H^2})(1+\frac{\delta\pi}{H^2})^2}\bigg].\nonumber
\end{eqnarray}

\begin{figure}[h!]
\centering
\includegraphics[scale=0.35]{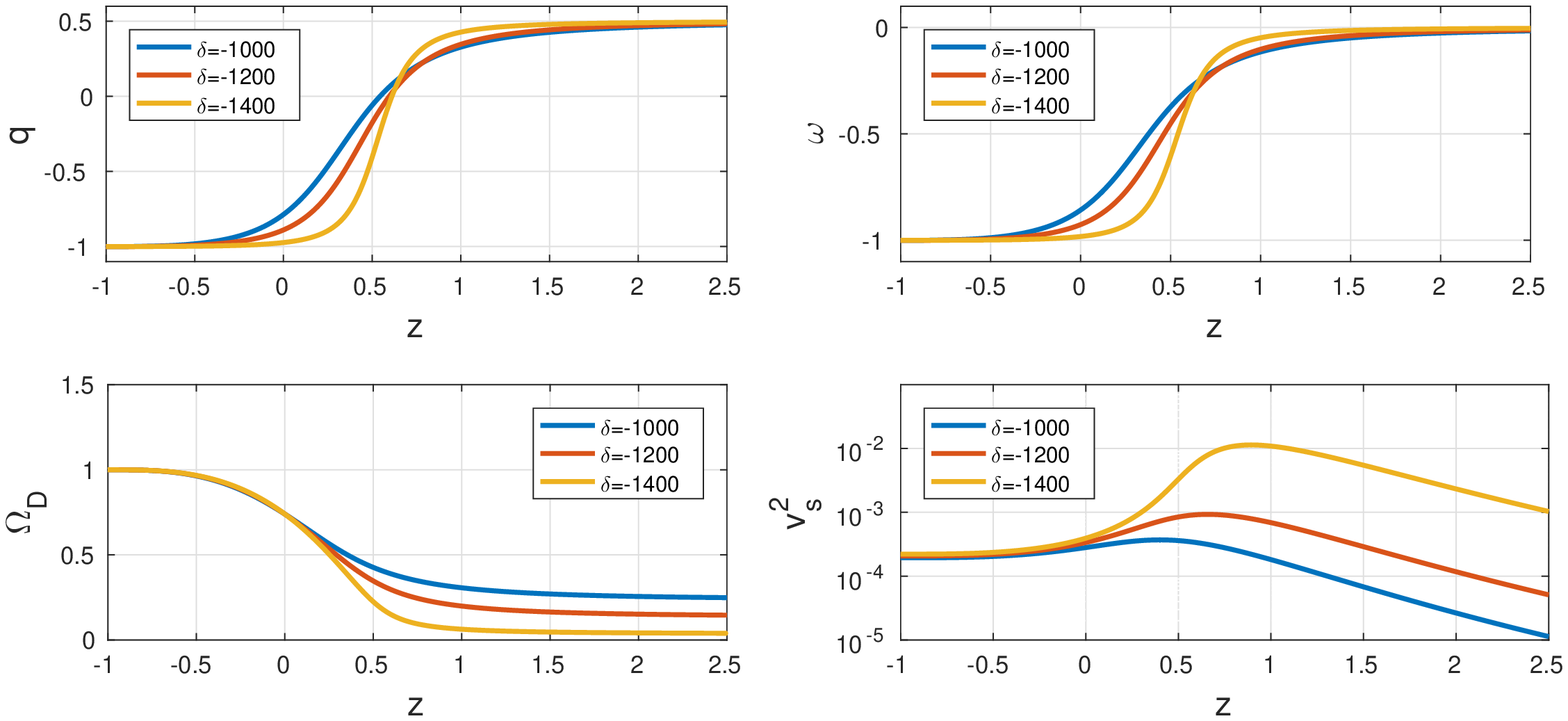}
\caption{$q$, $\Omega_D(z)$, $w$ and $v_{s}^{2}(z)$ versus $z$ for
some values of $\delta$. Here, $\Omega_m=0\cdot26$ and $H_0=67\
(Km/s)/Mpc$.\label{fig2}}
\end{figure}

\begin{figure}[h!]
\centering
\includegraphics[scale=0.36]{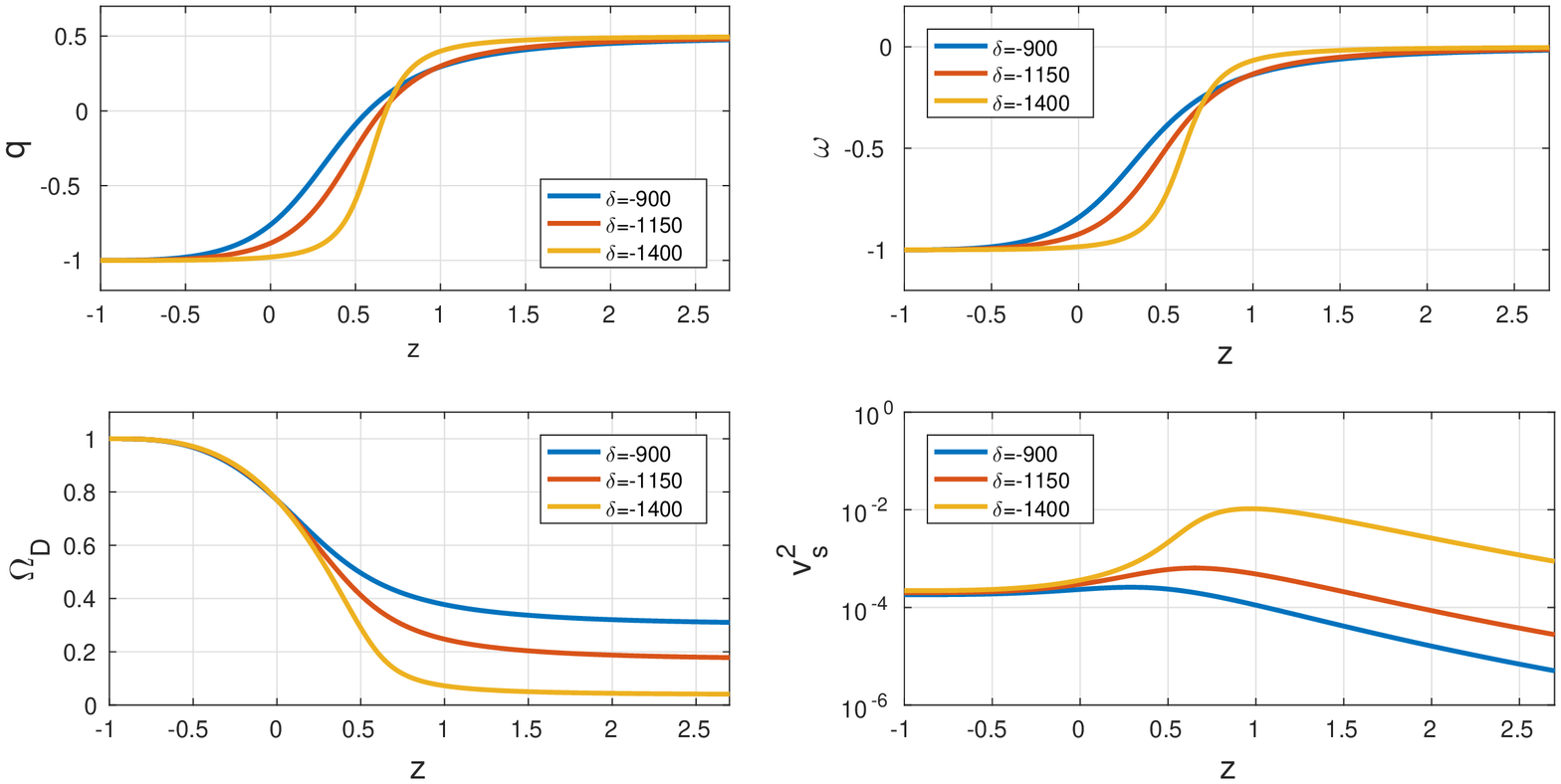}
\caption{$q$, $\Omega_D(z)$, $w$ and $v_{s}^{2}(z)$ versus $z$ for
some values of $\delta$. Here, $\Omega_m=0\cdot23$ and $H_0=67\
(Km/s)/Mpc$.\label{fig3}}
\end{figure}

In Figs.~(\ref{fig2}) and~(\ref{fig3}), the system parameters,
including $q$, $w$, $\Omega_D$ and $v_{s}^{2}$, have been plotted
versus $z$ for some values of $\delta$, whenever $\Omega_m=0\cdot26$
and $\Omega_m=0\cdot23$, respectively. In general, even for $z>z_t$,
$v_{s}^{2}$ can remain positive meaning that, unlike the OHDE and
SMHDE \cite{smm,stab}, RHDE can be stable in matter dominated era.
Moreover, at the high redshift limit, we have $q\rightarrow1/2$ and
$w\rightarrow0$, while at the $z\rightarrow-1$ limit,
$q\rightarrow-1$ and $w\rightarrow-1$. It is also worthwhile
mentioning that, depending on the value of $\delta$, the cosmos may
cross the phantom line ($q<-1$) for
$z<-1$. 

\begin{figure}[H]
\centering
\includegraphics[scale=0.35]{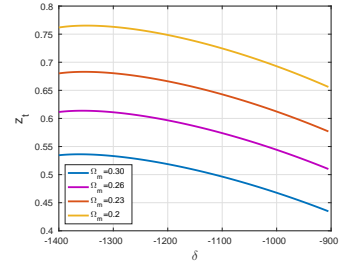}
\caption{Transition redshift ($z_t$) versus $\delta$ for some values
of $\Omega_m$ whenever $H_0=67\ (Km/s)/Mpc$.\label{fig1}}
\end{figure}

\noindent The transition redshift ($z_t$), at which
$q(z_t)=0$, versus $\delta$ has also been plotted in Fig.~(\ref{fig1})
for some values of $\Omega_m$ which lies within the
$0\cdot2\leq\Omega_m\leq0\cdot3$ range
\cite{Riess0,Riess,Riess1,mat1,mat2,mat3}. We see that the model can
give proper values for $z_t$, and as an example, the model predicts
that, depending on the value of $\delta$, $z_t\sim0\cdot5$ for
$\Omega_m=0\cdot3$ \cite{Riess0,Riess,Riess1,mat1,mat2,mat3}.

%
%
%
%
\section{Conclusion}

Recently, the notion of generalized entropy has been used to study
various properties of spacetime, gravitational and
cosmological phenomena. Here, by using FLT, we built a thermodynamic constraint on the relation between the system entropy
($S$) and the IR ($L$) and UV ($\Lambda$) cutoffs. Following this relation, using the R\'{e}nyi entropy, and considering the Bekenstein entropy
as the Tsallis entropy \cite{5,abe,non2,non19,non20,non21,non22}, we finally
proposed a new holographic model for dark energy (RHDE).

The model can generate acceptable values for the transition
redshift. We also studied the evolution of the system parameters
including $q$, $w$, $v_{s}^{2}$ and $\Omega_D$ which showed
satisfactory behavior by themselves. It has also been obtained that
RHDE shows more stability during the cosmic evolution
compared to SMHDE \cite{smm} and OHDE \cite{stab}.

\acknowledgments{We are grateful to the anonymous reviewer for
worthy hints and constructive comments. The work of H. Moradpour has
been supported financially by Research Institute for Astronomy \&
Astrophysics of Maragha (RIAAM) under project No. 1/5237-5. JPMG and
IPL are supported by CAPES (Coordena\c c\~ao de Aperfei\c coamento
de Pessoal de N\'ivel Superior).}


\begin{thebibliography}{99}
\bibitem{Riess0} P. Garnavich\textit{ et al}., Astrophys. J. 493, 53 (1998).
\bibitem{Riess} A. G. Riess\textit{ et al}., Astron. J. {\bf 116}, 1009 (1998).
\bibitem{Riess1} S. Perlmutter\textit{ et al}., Astrophys. J. {\bf 517}, 565 (1999).
\bibitem{Riess2} P. deBernardis, \textit{et al}., Nature {\bf 404}, 955 (2000).
\bibitem{Riess3} S. Perlmutter,\textit{et al}., Astrophys. J. {\bf 598}, 102 (2003).
\bibitem{COL2001} M. Colless \textit{et al}., Mon. Not. R. Astron. Soc. \textbf{328}, 1039 (2001).
\bibitem{COL20011} M. Tegmark \textit{etal}., Phys. Rev. D \textbf{69}, 103501 (2004).
\bibitem{COL20012} S. Cole\textit{ et al}., Mon. Not. R. Astron. Soc. \textbf{362}, 505 (2005).
\bibitem{COL20013} V. Springel, C. S. Frenk, S. M. D. White, Nature (London) \textbf{440}, 1137 (2006).
\bibitem{HAN2000} S. Hanany\textit{ et al}., Astrophys. J. Lett. \textbf{545}, L5 (2000).
\bibitem{HAN20001} C. B. Netterfield \textit{et al}., Astrophys. J. \textbf{571}, 604 (2002).
\bibitem{HAN20002} D. N. Spergel \textit{et al}., Astrophys. J. Suppl. \textbf{148}, 175 (2003).
\bibitem{mat1} Carl L. Gardner, Nucl. Phys. B 707, 278 (2005).
\bibitem{mat2} J. Ponce de Leon, Gen. Rel. Grav. 38, 61 (2006).
\bibitem{mat3} J. V. Cunha, Phys. Rev. D 79, 047301 (2009).
\bibitem{Roos} M. Roos, \textit{Introduction to Cosmology} (John Wiley and Sons, UK, 2003).
\bibitem{meeq} S. Capozziello, V. Faraoni, \textit{Beyond Einstein Gravity} (Springer, NY, 2011).
\bibitem{DEC3} C. Wetterich, Nucl. Phys. B {\bf302}, 668 (1988).
\bibitem{DEC6} R. R. Caldwell, M. Kamionkowski, N. N. Weinberg, Phys. Rev. Lett.{\bf 91}, 071301 (2003)
\bibitem{DEC8} C. Armendariz-Picon, V. F. Mukhanov, P. J. Steinhardt, Phys. Rev. Lett. {\bf85}, 4438 (2000)
\bibitem{DEC12} R. G. Cai, Phys. Lett. B {\bf657}, 228 (2007)
\bibitem{DEC13} H. Wei, R. G. Cai, Phys. Lett. B {\bf660}, 113 (2008)
\bibitem{DEC27} S. Nojiri, Sergei D. Odintsov, Phys. Rev. D {\bf72}, 023003 (2005).
\bibitem{DEC28} N. Ohta, Phys. Lett. B \textbf{695}, 41 (2011).
\bibitem{DEC372} H. Moradpour, A. Abri, H. Ebadi, Int. J. Mod. Phys. D 25, 1650014 (2016).
\bibitem{DEC39} H. Moradpour, R. C. Nunes, E. M. C. Abreu, J. A. Neto, Mod. Phys. Lett. A 32, 1750078 (2017).
\bibitem{Rev1} S. Nojiri, S. D. Odintsov, Phys. Lett. B {\bf639}, 144 (2006)
\bibitem{Rev2} K. Bamba, S. Capozziello, S. Nojiri, S. D. Odintsov, Astrophys. Space Sci. {\bf342}, 155 (2012).
\bibitem{mswr} H. Moradpour, A. Sheykhi, N. Riazi, B. Wang, AHEP. {\bf2014}, 718583 (2014).
\bibitem{em} H. Ebadi, H. Moradpour, Int. J. Mod. Phys. D {\bf24}, 1550098 (2015).
\bibitem{DEC37} H. Moradpour, M. T. Mohammadi Sabet, Can. J. Phys. {\bf94}, 1 (2016).
\bibitem{ijtpmr} H. Moradpour, N. Riazi, Int. J. Theor. Phys. 55, 268 (2016).
\bibitem{pav} J. P. Mimoso, D. Pav\'{o}n, Phys. Rev. D 94, 103507 (2016).
\bibitem{pla} M. Masi, Phys. Lett. A 338, 217 (2005).
\bibitem{nn1} H. Touchette, Physica A 305, 84 (2002).
\bibitem{nn2} T. S. Bir\'{o}, P. V\'{a}n, Phys. Rev. E 83, 061147 (2011).
\bibitem{nn3} C. Tsallis, Entropy 13, 1765 (2011).
\bibitem{non0} A. R\'{e}nyi, Probability Theory (North-Holland, Amsterdam, 1970).
\bibitem{non1} C. Tsallis, J. Stat. Phys. 52, 479 (1988).
\bibitem{abe} S. Abe, Phys. Rev. E 63, 061105 (2001).
\bibitem{fon} S. Abe, \textit{Foundations of Nonextensive Statistical Mechanics. In: Sengupta A. (eds) Chaos, Nonlinearity, Complexity. Studies in Fuzziness and Soft Computing} 206. (Springer, Berlin, Heidelberg (2006))
\bibitem{5} A. Majhi, Phys. Lett. B 775, 32 (2017).
\bibitem{non2} T. S. Bir\'{o}, V.G. Czinner, Phys. Lett. B 726, 861 (2013).
\bibitem{smm} A. Sayahian Jahromi et al., Phys. Lett. B 780, 21 (2018).
\bibitem{non16} A. Bialas, W. Czyz, EPL 83, 60009 (2008).
\bibitem{non22} V. G. Czinner, H. Iguchi, Phys. Lett. B 752, 306 (2016).
\bibitem{non19} N. Komatsu, Eur. Phys. J. C 77, 229 (2017).
\bibitem{non20} H. Moradpour, A. Bonilla, E. M. C. Abreu, J. A. Neto, Phys. Rev. D 96, 123504 (2017).
\bibitem{non21} H. Moradpour, A. Sheykhi, C. Corda, I. G. Salako, Phys. Lett. B 783, 82 (2018).
\bibitem{non13} H. Moradpour, Int. Jour. Theor. Phys. 55, 4176 (2016).
\bibitem{non4} E. M. C. Abreu, J. Ananias Neto, A. C. R. Mendes, W. Oliveira, Physica. A 392, 5154 (2013).
\bibitem{non5} E. M. C. Abreu, J. Ananias Neto. Phys. Lett. B 727, 524 (2013).
\bibitem{non6} E. M. Barboza Jr., R. C. Nunes, E. M. C. Abreu, J. A. Neto, Physica A: Statistical Mechanics and its Applications, 436, 301 (2015).
\bibitem{non7} R. C. Nunes, et al. JCAP, 08, 051 (2016).
\bibitem{HDE} A. G. Cohen, D. B. Kaplan, A. E. Nelson, Phys. Rev. Lett. {\bf82}, 4971 (1999)
\bibitem{HDE5} B. Guberina, R. Horvat, H. Nikoli\'{c}, JCAP 01, 012 (2007).
\bibitem{HDE17} S. Ghaffari, M. H. Dehghani, A. Sheykhi, Phys.  Rev. D {\bf89}, 123009 (2014).
\bibitem{HDE01} P. Horava, D. Minic, Phys. Rev. Lett. 85, 1610 (2000).
\bibitem{HDE1} S. Thomas, Phys. Rev. Lett. 89, 081301 (2002).
\bibitem{HDE2} S. D. H. Hsu, Phys. Lett. B {\bf594}, 13 (2004).
\bibitem{HDE3} M. Li, Phys. Lett. B {\bf603}, 1 (2004).
\bibitem{stab} Y. S. Myung, Phys. Lett. B 652, 223 (2007).
\bibitem{HDE6} W. Hao, Commun. Theor. Phys. 52, 743 (2009).
\bibitem{wang} B. Wang, E. Abdalla, F. Atrio-Barandela, D. Pavon, Rep. Prog. Phys. \textbf{79}, 096901 (2016).
\bibitem{RevH} S. Wang, Y. Wang, M. Li, Phys. Rep. 696, 1 (2017).
\bibitem{cons} S. A. Hayward, Class. Quantum Grav. {\bf 15}, 3147 (1998).
\bibitem{cons1} S. A. Hayward, S. Mukohyana, M. C. Ashworth, Phys. Lett.  A {\bf256}, 347 (1999)
\bibitem{cons2} D. Bak, S. J. Rey, Class. Quantum Grav. {\bf17}, 83 (2000).
\bibitem{cons3} R. G. Cai, S. P. Kim, J. High Energy Phys. 0502, 050 (2005).
\bibitem{Cai2} M. Akbar, R. G. Cai, Phys. Rev. D {\bf75}, 084003 (2007).
\bibitem{CaiKimt} R. G. Cai, L. M. Cao, Y. P. Hu, Class. Quantum. Grav. {\bf 26}, 155018 (2009).
\bibitem{my} Y. S. Myung, Phys. Lett. B 649, 247 (2007).
\bibitem{E1} S. Capozziello, O. Luongo, arXiv:1704.00195v1.
\bibitem{HDET} O. Luongo, AHEP, 2017, Article ID 1424503 (2017).
\bibitem{vv1} E. Chang-Young, D. Lee, JHEP 1404, 125 (2014).
\bibitem{vv2} M. Eune, W. Kim, Phys. Rev. D 88, 067303 (2013).
\bibitem{j1} T. Jacobson, Phys. Rev. Lett. 75, 1260 (1995).
\bibitem{j2} A. V. Frolov, L. Kofman, JCAP 0305, 009 (2003).
\end{thebibliography}
\end{document}